\documentclass{article}

\usepackage{PRIMEarxiv}

\usepackage[utf8]{inputenc} % allow utf-8 input
\usepackage[T1]{fontenc}    % use 8-bit T1 fonts
\usepackage{underscore} % allows breaks at underscores
\usepackage{hyperref}       % hyperlinks
\usepackage{url}            % simple URL typesetting
\usepackage{booktabs}       % professional-quality tables
\usepackage{amsfonts}       % blackboard math symbols
\usepackage{nicefrac}       % compact symbols for 1/2, etc.
\usepackage{microtype}      % microtypography
\usepackage{lipsum}
\usepackage{fancyhdr}       % header
\usepackage{graphicx}       % graphics
\graphicspath{{media/}}     % organize your images and other figures under media/ folder
\usepackage{float}

\title{TLS Certificate and Domain Feature Analysis of Phishing Domains in the Danish .dk Namespace
}

\author{
\normalfont
\textbf{Athanasios P. Pelekoudas\textsuperscript{*},
Epameinondas Bolis\textsuperscript{*},
Jasmin Lindner\textsuperscript{*},
Prodromos Kyriakidis\textsuperscript{*}},\\
\textbf{Mathias Davidsen\textsuperscript{*},
Johannes T. E. Hansen\textsuperscript{*},
Christian H. Reichkendler\textsuperscript{*},
Sajad Homayoun\textsuperscript{**}}\\[4pt]
Aalborg University, Copenhagen, Denmark\\[4pt]
\texttt{\{apelek25, ebolis25, jlindn25, pkiria25, mdav21, jteh21, creich21\}@student.aau.dk}\\
\texttt{sajadh@es.aau.dk}
}

\begin{document}
\maketitle
\vspace{-0.8cm}
\begin{center}
\footnotesize\textit{\textsuperscript{*}These authors contributed equally to this work.}\\
\footnotesize\textit{\textsuperscript{**}Corresponding author: sajadh@es.aau.dk}
\end{center}
\vspace{0.5cm}

\begin{abstract}
Phishing attacks remain a persistent cybersecurity threat, and the widespread adoption of TLS certificates has unintentionally enabled malicious websites to appear trustworthy to users. This study examines whether certificate metadata and domain characteristics can help distinguish phishing domains from benign domains within the Danish .dk namespace. A dataset was constructed by combining registry information from Punktum dk with phishing reports and popularity rankings from external sources. TLS certificate attributes were collected using Netlas, while additional domain-based features were derived from DNS records and lexical analysis of domain names. The analysis compares phishing, popular, and less frequently visited domains across several feature categories, including Certificate Authorities (CAs), validity periods, missing certificate fields, SAN structure, registrant geography, hosting providers, and lexical properties of domain names. The results indicate that several features show observable differences between phishing and highly popular domains. However, phishing domains often resemble less popular domains, resulting in substantial overlap across many characteristics. Consequently, no individual feature provides a reliable standalone indicator of phishing activity within the Danish namespace. The findings suggest that certificate and domain attributes may still contribute to detection when combined, while also highlighting the limitations of relying on individual indicators in isolation. This work provides an empirical overview of phishing-related infrastructure patterns in the Danish .dk ecosystem and offers insights that may inform future phishing detection approaches. 
\end{abstract}

% keywords can be removed

\section{Introduction}

TLS certificates play a central role in establishing trust on the web. However, attackers increasingly exploit this trust model by obtaining legitimate certificates for malicious domains. Adversaries exploit these mechanisms to deploy phishing campaigns targeting unsuspecting users. Phishing attacks represent a persistent threat to cybersecurity, as attackers leverage social engineering techniques to deceive users into divulging confidential credentials and other sensitive information ~\cite{nirmal2023EffCer,Bijmans20213757}. To enhance the perceived legitimacy of their malicious activities and evade security controls, threat actors increasingly utilize fraudulent domains and valid TLS certificates that imitate trusted or reputable organizations ~\cite{Ishida2023174,LiJiaxin2021MLMX}.

Transport Layer Security (TLS) certificates are used to authenticate servers and establish encrypted communication channels between clients and websites ~\cite{secrypt21}. These certificates provide cryptographic protection and a browser-recognized indication of domain control, characteristics that users have been trained to associate with legitimacy and trustworthiness ~\cite{nirmal2023EffCer}. When attackers obtain valid certificates for malicious domains, phishing websites can display the browser security padlock while impersonating legitimate organizations. This reduces user suspicion and increases the likelihood of successful credential theft ~\cite{LiJiaxin2021MLMX}.

Certificates are issued and signed by trusted third-party entities known as CAs, which form a hierarchical trust model relied upon by modern browsers ~\cite{OppligerRolf2016SaTt}. While this model is designed to enable secure authentication, prior studies show that attackers can exploit it by obtaining certificates through automated or weak validation processes, allowing malicious domains to appear legitimate ~\cite{SakuraiIdePh2021}. To increase transparency and accountability in certificate issuance, Certificate Transparency (CT) logs record all newly issued certificates in publicly auditable, append-only logs ~\cite{secrypt21}. CT data has therefore become a valuable resource for studying large-scale domain registration patterns, identifying fraudulent certificates, and analyzing phishing infrastructure [~\cite{OppligerRolf2016SaTt}~\cite{hranickyUnmPh2024} .

In response to the growing abuse of certificates in phishing operations, certificate analysis has emerged as an important methodological approach for understanding phishing infrastructure ~\cite{nirmal2023EffCer}. TLS certificates contain substantial metadata, including issuer information, validity periods, SANs, and cryptographic parameters, which can reveal patterns related to certificate issuance, infrastructure reuse, and attack campaign deployment ~\cite{secrypt21}. Through systematic examination of these attributes, suspicious patterns can be detected, related phishing campaigns can be linked, and infrastructure-level detection rules can be developed ~\cite{SakuraiIdePh2021}. Nevertheless, existing research also highlights that no single certificate feature is sufficient to reliably distinguish phishing from benign domains, as many characteristics previously associated with phishing have become common among legitimate websites.

Although considerable research has addressed certificate-based phishing analysis, most existing studies focus on global datasets and common top-level domains such as .com, .net, and .org, leaving country-code top-level domains substantially understudied ~\cite{quinkertDigDe2021}. In particular, the Danish .dk namespace has not been adequately examined using certificate analysis, despite its exposure to phishing activity and its distinct operational characteristics related to domain registration practices, certificate authority usage patterns, and registrant behavior ~\cite{secrypt21}.

To address this gap, this study conducts a quantitative, hypothesis-driven analysis of TLS certificates used by phishing and legitimate domains within the Danish namespace. In collaboration with Punktum dk A/S \footnote{\url{https://punktum.dk/}}, the administrator of the .dk domain registry, datasets of phishing domains and benign domains are collected and analyzed across multiple feature categories, including certificate properties, domain characteristics, registrant information, and sectoral targeting patterns. Building on insights from prior work, the study evaluates a set of hypotheses concerning certificate validity duration, issuing authorities, field completeness, SAN structure, registrant geography, and impersonated sectors. Rather than proposing a standalone detection system, the goal is to characterize phishing infrastructure behavior within a national domain context and assess the strengths and limitations of certificate-based indicators for ccTLD ecosystems.

\textbf{This paper makes the following contributions:}

\begin{itemize}
    \item We construct a dataset of phishing and benign domains within the Danish \texttt{.dk} namespace by combining registry data from Punktum dk, phishing reports from AbuseManager, and domain popularity information from the Tranco list.
    
    \item We perform a quantitative analysis of TLS certificate metadata and domain-based features across phishing, popular, and less frequently used domains.
    
    \item We evaluate the effectiveness of commonly used certificate and domain indicators for distinguishing phishing domains in a country-code top-level domain (ccTLD) ecosystem.
    
    \item We provide insights into sectoral targeting patterns of phishing domains in Denmark, identifying which types of organizations are most frequently impersonated in phishing campaigns.
\end{itemize}

The remainder of this paper is organized as follows. Section~\ref{sec:related_work} reviews related work on phishing detection using TLS certificates and domain-based features. Section~\ref{sec:methodoly} describes the data sources and methodology. Section~\ref{sec:results} presents the analysis and results. Section~\ref{sec:discussion} discusses limitations and future work, and Section~\ref{sec:conclusion} concludes the paper.

\section{Related Work}
\label{sec:related_work}
\subsection{Phishing Detection Using TLS Certificate Metadata}

Prior research has explored a variety of approaches for detecting phishing domains and malicious TLS certificates. These approaches typically analyze certificate metadata, domain characteristics, and DNS information using machine learning models, heuristic detection rules, or graph-based analysis of relationships between domains and certificates.

Phishing detection is commonly formulated as a binary classification task that distinguishes malicious domains from benign ones. Several studies apply supervised learning models such as Random Forests, XGBoost, and logistic regression, often combining certificate metadata with domain-based and WHOIS-derived features.

More recent studies explore neural and graph-based approaches that capture structural relationships between domains and certificates. Liu et al.~\cite{Liu2022} propose a Graph Convolutional Network that models certificate reuse and SAN structures, while Shashwat et al.~\cite{Shashwat2024173} use sentence embeddings to represent issuer and subject fields as semantic vectors that can be compared using similarity measures.

Unsupervised techniques are also used to detect anomalies without labeled data.
Ishida et al. \cite{Ishida2023174} apply graph-based clustering over DNS and certificate data,
extracting features such as SAN overlaps and certificate reuse. AlSabah et al.
\cite{AlSabahMashael2022CDoP} incorporate CT log and passive DNS signals, including issuance timing,
TTL variability, and IP rotation.

Heuristic-based approaches remain relevant as complementary tools that rely on manually defined rules to identify suspicious certificate and domain properties. For example, Hageman et al.~\cite{secrypt21} highlight indicators such as short certificate lifetimes, exclusive reliance on DV certificates, and the absence of DNSSEC. Similarly, Kim et al.~\cite{Kim2021407} incorporate lexical domain patterns and WHOIS privacy indicators.

\subsection{Feature Extraction}
In addition to detection techniques, prior research has focused on identifying informative features extracted from TLS certificates, domain names, and DNS infrastructure. Feature extraction plays a central role in certificate-based phishing detection by transforming raw TLS certificate fields and associated domain metadata into structured indicators. Prior work commonly groups extracted features into certificate-based, SAN-based, domain-based, WHOIS-based, and CT log-derived
categories, each providing complementary signals of trust and malicious behavior.

\subsubsection{Certificate and SAN Features}

Certificate-based features are extracted directly from the X.509 standard,
including subject and issuer attributes such as the Common Name, Organization,
and Country. Haraldsdóttir et al. \cite{haraldsdottirUnPh2024} note that inconsistencies or generic
placeholders in these fields often correlate with malicious intent. Validation level
is another key indicator. DV certificates are frequently abused due to minimal
issuance requirements, whereas OV and EV certificates demand stricter identity
verification \cite{Kim2021407}. The certificate validity period, derived from the
\textit{notBefore} and \textit{notAfter} fields, is also widely used, as phishing
domains often rely on short-lived certificates to reduce exposure time \cite{LiJiaxin2021MLMX}.

SAN-based features capture the structure of the Subject Alternative Name
extension, including the number of covered hostnames, their diversity, entropy,
and wildcard usage. High SAN counts may indicate certificate reuse across many
domains, which is commonly associated with phishing infrastructure. Liu et al.
\cite{Liu2022} and Shashwat et al. \cite{Shashwat2024173} extract SAN statistics such as entry counts
and list lengths, while other studies examine SAN similarity patterns to detect
automatically generated certificates.

\subsubsection{Domain-Based Features}

Beyond certificate fields, phishing detection systems incorporate lexical and
structural domain characteristics. Malicious domains often appear longer, include
multiple subdomains, and exhibit higher entropy values \cite{hranickyUnmPh2024}. Similarity
metrics such as Levenshtein distance are frequently applied to identify
typosquatting and brand impersonation attempts.

WHOIS-based indicators provide additional context on registration history,
including registrar identity, creation dates, and overall domain age. Domain age
is consistently cited as one of the strongest predictors, as attackers often register domains shortly before deploying them in phishing campaigns \cite{Erfan202414}. CT logs provide further time-based signals, where gaps between certificate issuance and DNS activation may reveal coordinated or automated attacker behavior \cite{nirmal2023EffCer}.

\subsection{Feature Categories in Prior Work}

Certificate-based attributes remain the most frequently used feature category in
prior phishing detection studies, reflecting the central role of TLS certificate
metadata in malicious domain classification. Beyond standard X.509 fields,
researchers commonly extract complementary indicators from SAN structures and
domain-level lexical or registration characteristics.

Table~\ref{tab:cert_features} summarizes the most common groups of
certificate-based indicators, including subject and issuer information, validation
properties, and cryptographic parameters. In addition, SAN-derived attributes are
widely used to capture certificate reuse patterns and hostname diversity, as shown
in Table~\ref{tab:san_features}. Domain-based feature categories frequently
leveraged for phishing classification, including lexical structure and WHOIS-based
metadata, are presented in Table~\ref{tab:domain_features}.

\begin{table}[h]
    \centering
    \caption{Certificate-Based Feature Grouping}
    \renewcommand{\arraystretch}{1.15}
    \begin{tabular}{lp{5cm}cp{3cm}}
        \toprule
        \textbf{Feature Group} &
        \textbf{Most used features} &
        \textbf{Paper Count} &
        \textbf{Papers} \\
        \midrule
        Subject Information &
        subject name, Common Name, organization, email &
        10 &
        \cite{haraldsdottirUnPh2024,Shashwat2024173,Alkinoon2023,Liu2022,Kim2021407,LiJiaxin2021MLMX,quinkertDigDe2021,SakuraiIdePh2021,Bijmans20213757,nirmal2023EffCer} \\

        Validation \& Trust Indicators &
        validation\_level, expiration, validity period &
        7 &
        \cite{haraldsdottirUnPh2024,hranickyUnmPh2024,Alkinoon2023,AlSabahMashael2022CDoP,Liu2022,LiJiaxin2021MLMX,secrypt21} \\

        Issuer Information &
        issuer name, authority info, policies &
        7 &
        \cite{haraldsdottirUnPh2024,Alkinoon2023,AlSabahMashael2022CDoP,Liu2022,Kim2021407,LiJiaxin2021MLMX,secrypt21} \\

        Key \& Algorithm Properties &
        public key algorithm, key size, usage &
        6 &
        \cite{haraldsdottirUnPh2024,Alkinoon2023,AlSabahMashael2022CDoP,Liu2022,Kim2021407,LiJiaxin2021MLMX} \\

        Extended Policy \& Extensions &
        cert\_policies, constraints, authority\_id &
        4 &
        \cite{haraldsdottirUnPh2024,hranickyUnmPh2024,AlSabahMashael2022CDoP,Liu2022} \\

        Chain, OCSP \& CRL Data &
        ocsp\_urls, crl\_dist\_point\_present &
        3 &
        \cite{haraldsdottirUnPh2024,AlSabahMashael2022CDoP,Liu2022} \\

        Error \& Verification Indicators &
        error\_occur, parse\_error, verify\_error &
        2 &
        \cite{Liu2022,LiJiaxin2021MLMX} \\
        \bottomrule
    \end{tabular}
    \label{tab:cert_features}
\end{table}

\begin{table}[h]
\centering
\caption{SAN-Based Feature Grouping}
\renewcommand{\arraystretch}{1.2}
\begin{tabular}{lp{5cm}cp{2.3cm}}
\toprule
\textbf{Feature Group} &
\textbf{Most used features} &
\textbf{Paper Count} &
\textbf{Papers} \\
\midrule
SAN Size \& Statistical Metrics &
count/list of SANs, mean\_san\_domain\_len, num\_tokens &
6 &
\cite{haraldsdottirUnPh2024,hranickyUnmPh2024,AlSabahMashael2022CDoP,Liu2022,Kim2021407,secrypt21} \\

Suspiciousness \& Lexical Complexity &
san\_sus\_keyword, san\_tld, entropy, char\_diversity &
3 &
\cite{haraldsdottirUnPh2024,hranickyUnmPh2024,AlSabahMashael2022CDoP} \\

Structural Indicators &
san\_token\_is\_tld, san\_is\_international, lcs\_sans &
2 &
\cite{haraldsdottirUnPh2024,hranickyUnmPh2024} \\

Wildcard \& Ownership Heuristics &
wildcard detection, SAN--subject match &
2 &
\cite{haraldsdottirUnPh2024,Erfan202414} \\
\bottomrule
\end{tabular}

\label{tab:san_features}
\end{table}

\begin{table}[H]
\centering
\caption{Domain-Based Feature Grouping}
\renewcommand{\arraystretch}{1.2}
\begin{tabular}{lp{5cm}cp{3cm}}
\toprule
\textbf{Feature Group} &
\textbf{Most used features} &
\textbf{Paper Count} &
\textbf{Papers} \\
\midrule
Lexical \& Structural &
domain\_len, num\_tokens, hyphens, subdomain count &
6 &
\cite{haraldsdottirUnPh2024,hranickyUnmPh2024,Erfan202414,AlSabahMashael2022CDoP,Bijmans20213757,nirmal2023EffCer} \\

Suspicious Pattern Indicators &
sus\_keyword, sus\_tld, typosquatted, brand names &
6 &
\cite{haraldsdottirUnPh2024,hranickyUnmPh2024,Erfan202414,quinkertDigDe2021,Bijmans20213757,nirmal2023EffCer} \\

Registration \& WHOIS Metadata &
domain creation date, age, registrar info &
4 &
\cite{dogan2024DuLa,hranickyUnmPh2024,Alkinoon2023,nirmal2023EffCer} \\

Entropy Metrics &
shannon\_entropy &
3 &
\cite{haraldsdottirUnPh2024,hranickyUnmPh2024,AlSabahMashael2022CDoP} \\
\bottomrule
\end{tabular}

\label{tab:domain_features}
\end{table}

Although phishing detection using TLS certificate analysis has been extensively studied on global datasets and common top-level domains such as \texttt{.com} or \texttt{.org}, country-code namespaces such as \texttt{.dk} remain largely underexplored. To the best of our knowledge, no prior work has systematically analyzed TLS certificate characteristics associated with phishing domains within the Danish namespace. This motivates a focused empirical analysis of certificate and domain features observed in phishing domains targeting the Danish \texttt{.dk} ecosystem.

To address this gap, this paper investigates whether certificate metadata and domain-based features can provide meaningful signals for distinguishing phishing domains from benign domains within the Danish \texttt{.dk} namespace. The study focuses on empirical analysis of certificate and domain characteristics observed in real-world phishing activity targeting Danish domains.

\section{Methodology and Data Collection}
The following section covers the approach in collecting and processing data. This includes data sources, pipeline and feature enrichment.
\label{sec:methodoly}

\subsection{Data sources}
To cover features most commonly found across literature, four different sources of data are examined.
The first and most important dataset is provided by Punktum dk \footnote{\url{https://punktum.dk/}} and contains a list of all registered domains in the .dk namespace, along with their registrant country, validation status and an anonymized user Id.

The second data source is also provided by Punktum dk but originates from Abusemanager \footnote{Abusemanager is a threat intelligence platform which monitors domain abuse. \url{https://iq.global/iq-abuse-manager}} and retrieved on the 6th of November 2025. This dataset is a blocklist of malicious sites in the Danish namespace, which will be used to identify phishing domains. Because Punktum dk themselves only recently acquired access, this data only goes back to the beginning of 2024. %, but this will be addressed later in section \ref{sec:limitations}. 
The list includes domains reported as spam, phishing, and malware.
Since the focus is on phishing domains, the data is filtered accordingly.

The third dataset is the publicly available Tranco top 1 million list \footnote{Tranco 1 Million list is a ranking of the most popular websites. \url{https://tranco-list.eu/}} retrieved on the 10th of November 2025. Since Tranco serves as a ranking of the most visited websites, domains listed are considered highly likely to be legitimate. By comparing this list with the registrant list, popular danish domains can be identified. This is done under the assumption that popular domains may behave very differently from unpopular domains.

Lastly, any domain not found in either the AbuseManager blocklist or the Tranco list is assigned the label \textit{unpopular}. These domains are treated as likely benign for the purpose of this analysis, although it cannot be guaranteed that all such domains are truly benign. This labeling approach follows common practice in phishing measurement studies, where domains not present in abuse datasets are treated as benign proxies for comparison.

\subsection{Pipeline}
To extract useful information from multiple sources, the first step is to process the data and enrich it across data points. The domain name is used as a common identifier across the data, as this is the feature most prevalent across data. 

\begin{figure}[h]
  \centering
  \includegraphics[width=0.9\linewidth]{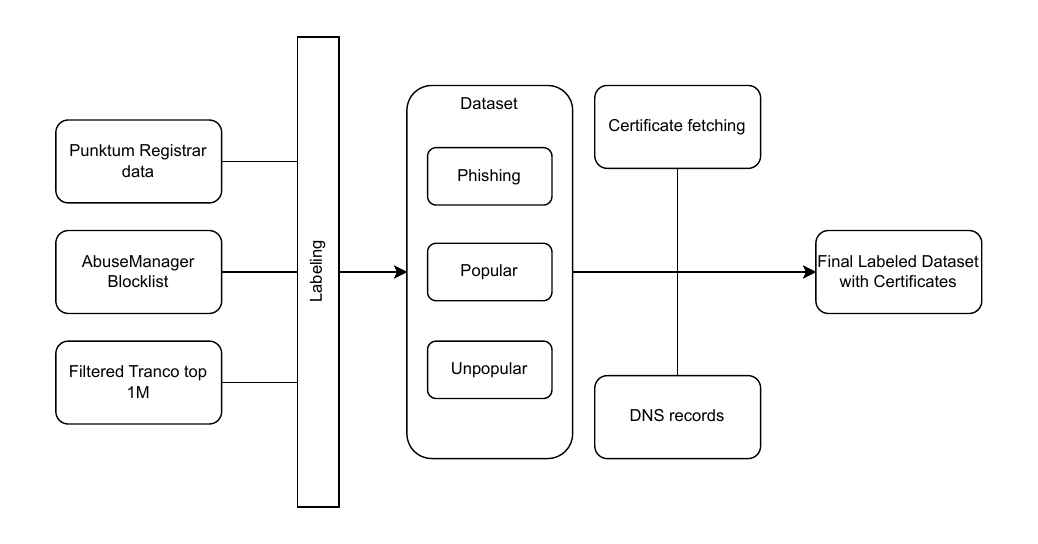}
  \caption{Pipeline Diagram}
  \label{fig:Pipeline}
\end{figure}

The pipeline as shown in figure \ref{fig:Pipeline} is structured around three phases: labeling, enrichment, and filtering.

To create a labeled dataset, the registrar dataset was enriched using two external sources. First the dataset was compared with the filtered blocklist using the domain name as the key. If a domain appeared in both datasets, it was assigned the label \textit{phishing}. This resulted in 762 unique domains. This step separated a small but highly relevant subset of phishing domains from the large pool of unlabeled .dk entries.

Additionally, the domains that appeared in the Tranco list were labeled as popular. After filtering the list to only include .dk domains, the list was compared with the primary dataset again based on the domain name. This matching resulted in 2135 domains. These domains represent high-reputation, frequently visited websites that are unlikely to be phishing.

\subsection{Feature Enrichment}
After labeling the primary dataset, different certificate-based features were extracted using Netlas.io \footnote{Netlas is an search engine that collects and indexes data such as domains, IP addresses, certificates, open ports, and web services. \url{https://netlas.io/}}. These features include, but are not limited to, temporal attributes, subject and issuer information, algorithm information, and SAN metadata. 

Due to the time required to retrieve certificate data from the Netlas platform, certificate information was collected for a limited subset of domains from each category. The final dataset includes 762 phishing domains, 1260 popular domains, and 2000 unpopular domains.

\begin{table}[h]
 \caption{Dataset summary by label}
  \centering
  \begin{tabular}{lllll}
    \toprule
    Label & Total domains & With certificates & Coverage & Certificate count\\
    \midrule
    Phishing & 762 & 270 & 35.4\% & 4668\\
    Popular & 1260 & 1110 & 88.1\% &  8934\\
    Unpopular & 2000 & 825 & 42.6\% & 15212 \\
    \bottomrule
  \end{tabular}
  \label{tab:cert_cov}
\end{table}

Not every domain in the dataset had an available certificate at the time of collection, as shown in table \ref{tab:cert_cov} in the column Coverage. 
It was observed that a substantial proportion of domains in both the Phishing (64.6\%) and Unpopular (57.4\%) categories lacked a certificate, whereas in the Popular domains only 11.9\% were missing a certificate. While this discrepancy could potentially indicate differences in deployment practices, it is also possible that the Netlas API was unable to retrieve certain certificates from its internal dataset. Netlas does not perform live queries against CT logs and instead relies on its internally maintained dataset of observed certificates. As a result, some recently issued or less frequently observed certificates may not appear in the search results. This limitation may also explain why some popular domains were missing certificates in the collected dataset.

Table \ref{tab:cert_cov} also shows the certificate count for every label which illustrate that many domains have more than one certificate. To ensure consistency in the analysis, only one certificate was selected for each domain. Including all observed certificates would bias the dataset toward domains with frequent certificate renewals or multiple certificate deployments. Therefore, for popular and unpopular domains the most recent certificate was selected to represent the current configuration of the domain. For phishing domains, the certificate closest to the AbuseManager report date was selected in order to represent the configuration of the domain at the time it was identified as malicious.

To complete the certificate information, several other domain-based features were also retrieved using DNS records. They capture registration behavior as well as structural and temporal information about a given domain.

\section{Results and Discussion}
\label{sec:results}
This section presents the results of the analysis of certificate- and domain-related characteristics associated with phishing activity in the Danish domain space. The analysis used the aforementioned dataset.

\subsection{Issuer organization}
In this hypothesis, the different CAs within the dataset of phishing domains and their associated certificates were examined and correlated, with the CA information in popular and unpopular. The analysis primarily focuses on the field "issuer\_organization" which indicates which CA issued the certificate. After analyzing the CAs pricing, it was observed that 5 out of 11 providers (45\%) offer a free plan as seen on Table \ref{tab:CA_pricing}. While the remainders provide some form of free trial or refund.

\begin{table}[h]
    \centering
    \caption{CA's pricing options}
    \begin{tabular}{ll}
    \toprule
    CAs&Pricing\\ \midrule
    Amazon & Free/Paid \\
    Cloudflare & Free/Paid \\
    Comodo & Paid\\
    cPanel & Free Trial/Paid\\
    DigiCert & Refund option/Paid\\
    GoDaddy & Refund option/Paid\\
    Google Trust Services & Free/Paid\\
    Let's Encrypt & Free\\
    Sectigo & Refund option/Paid\\
    Sonera & Paid\\
    ZeroSSL &Free/Paid\\
    \bottomrule
    \end{tabular}
    \label{tab:CA_pricing}
\end{table}

\begin{figure}[h]
    \centering
    \includegraphics[width=0.8\linewidth]{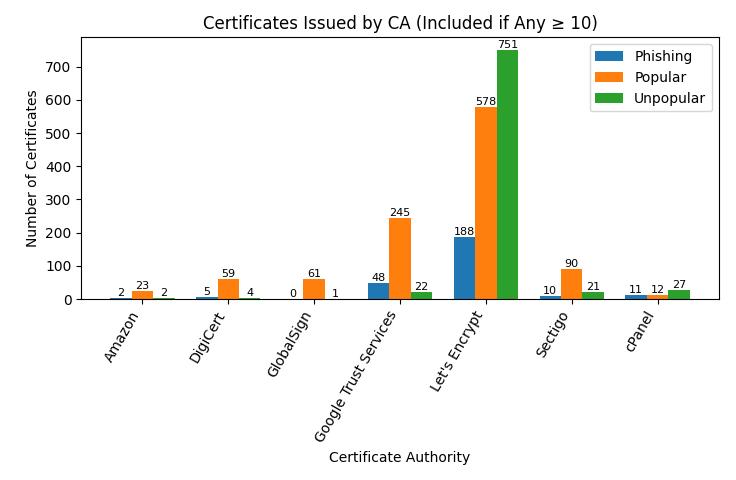}
    \caption{Number of issued certificates by CA}
    \label{fig:Issued_certs}
\end{figure}

Figure \ref{fig:Issued_certs} shows the frequency of different CAs in the phishing dataset. Google Trust Services and Let’s Encrypt are the dominant issuers, together accounting for 87.4\% of observed certificates in phishing. As shown in Table \ref{tab:CA_pricing}, both offer free certificates, supporting the hypothesis that \textit{"A high number of phishing certificates are issued by a small set of free CAs"}. Let’s Encrypt and Google Trust Services account for 74.6\% of certificates in the popular dataset, while Let’s Encrypt alone issues 87.6\% of certificates in the unpopular dataset. These results closely resemble the results of the phishing dataset.

Overall, unpopular and phishing domains are nearly indistinguishable when comparing for this feature, while popular domains show only minor deviation. The differences between categories are insufficient to reliably classify domains as phishing or benign. Therefore, even though the hypothesis is true, this feature does not have a strong indicator to determine whether a certificate is registered for phishing.

\subsection{Validity Period}

The hypothesis analyzed here is that certificates issued to domains that were registered with the intent of conducting phishing activity will have a shorter validity period than domains registered to domains without the intent of phishing. To test this hypothesis, all 3 datasets need to be analyzed. First, the validity durations in the datasets will be put into distributions, which will be used to better understand the mean, median, and mode of each dataset. Table \ref{tab:ttl_averages} below shows the means, medians, and modes of validity lengths within each dataset. 

\begin{table}[H]
    \centering
    \caption{Table showing the mean, median, and mode of certificates' validity durations in each dataset. All values are in seconds}
    \begin{tabular}{llll}
        \toprule
         & Mean & Median & Mode \\ \midrule
        Phishing & 9,395,025.19 & 7,775,999.00 & 7,775,999.00 \\
        Popular & 15,442,399.46 & 7,775,999.00 & 7,775,999.00 \\
        Unpopular & 9,309,925.38 & 7,775,999.00 & 7,775,999.00 \\
        \bottomrule
    \end{tabular}
    \label{tab:ttl_averages}
\end{table}

\begin{figure}[h]
    \centering
    \includegraphics[width=1\linewidth]{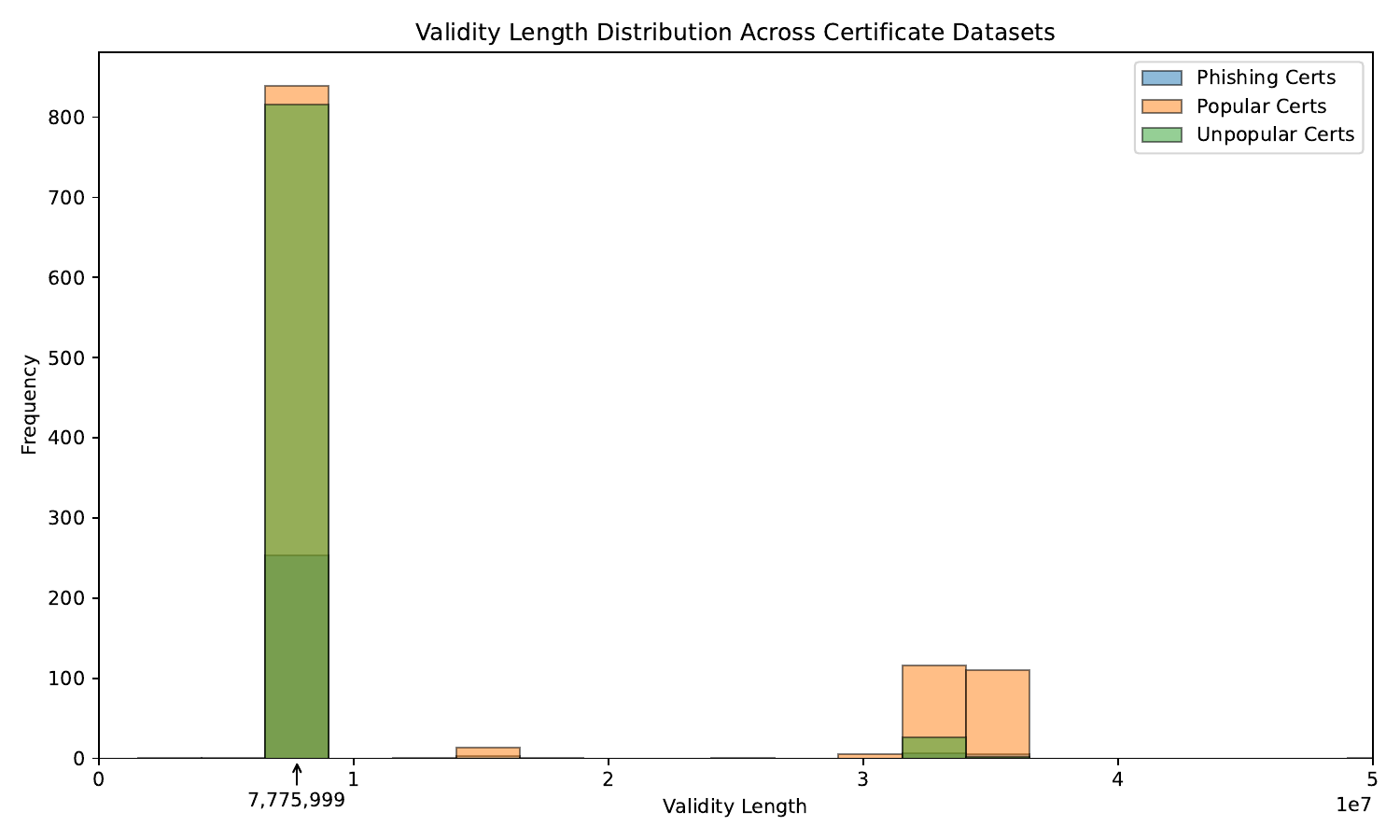}
    \caption{Histogram showing the total frequency of certificate lifetimes for all datasets}
    \label{fig:histogram_total_freq}
\end{figure}

Across all 3 datasets, the medians and modes are the same. This value is \textit{7,775,999} which, when converted from seconds to months, corresponds to almost 3 months. This overrepresentation might be due to the popularity of Let's Encrypt. As of collecting certificates, Let's Encrypt only issued certificates with validity lengths of 90 days and in some cases 6 days. The mean validity duration of certificates from phishing and unpopular domains are almost indistinguishable from each other. The mean for certificates from popular domains is almost twice the length of the median and mode values. This means that while the mode and median is 3 months, popular domains do have a bigger tendency to register certificates with longer validity duration, thus shifting the mean up.
Figure \ref{fig:histogram_total_freq} below is a histogram showing the validity lengths of certificates in all datasets. The bin width is 2,500,000. 

The figure shows that a large majority of all certificates have a validity length between $\sim$6,500,000 and $\sim$9,000,000 seconds. In total, this bin contained 1908 certificates, with 253 of them being from phishing, 839 from popular, and 816 from unpopular. Within the popular domains, there was a spike of certificates with lifetimes of around 34,000,000 seconds, corresponding to around 13 months. Specifically, 226 certificates landed in the two bins surrounding this value. \\
While this figure provides insight to the total amount of certificates, there is an imbalance in the amount of certificates between the datasets. To better understand these numbers, a histogram with relative frequency can be seen in Figure \ref{fig:histogram_relative_freq} below.

\begin{figure}[h]
    \centering
    \includegraphics[width=1\linewidth]{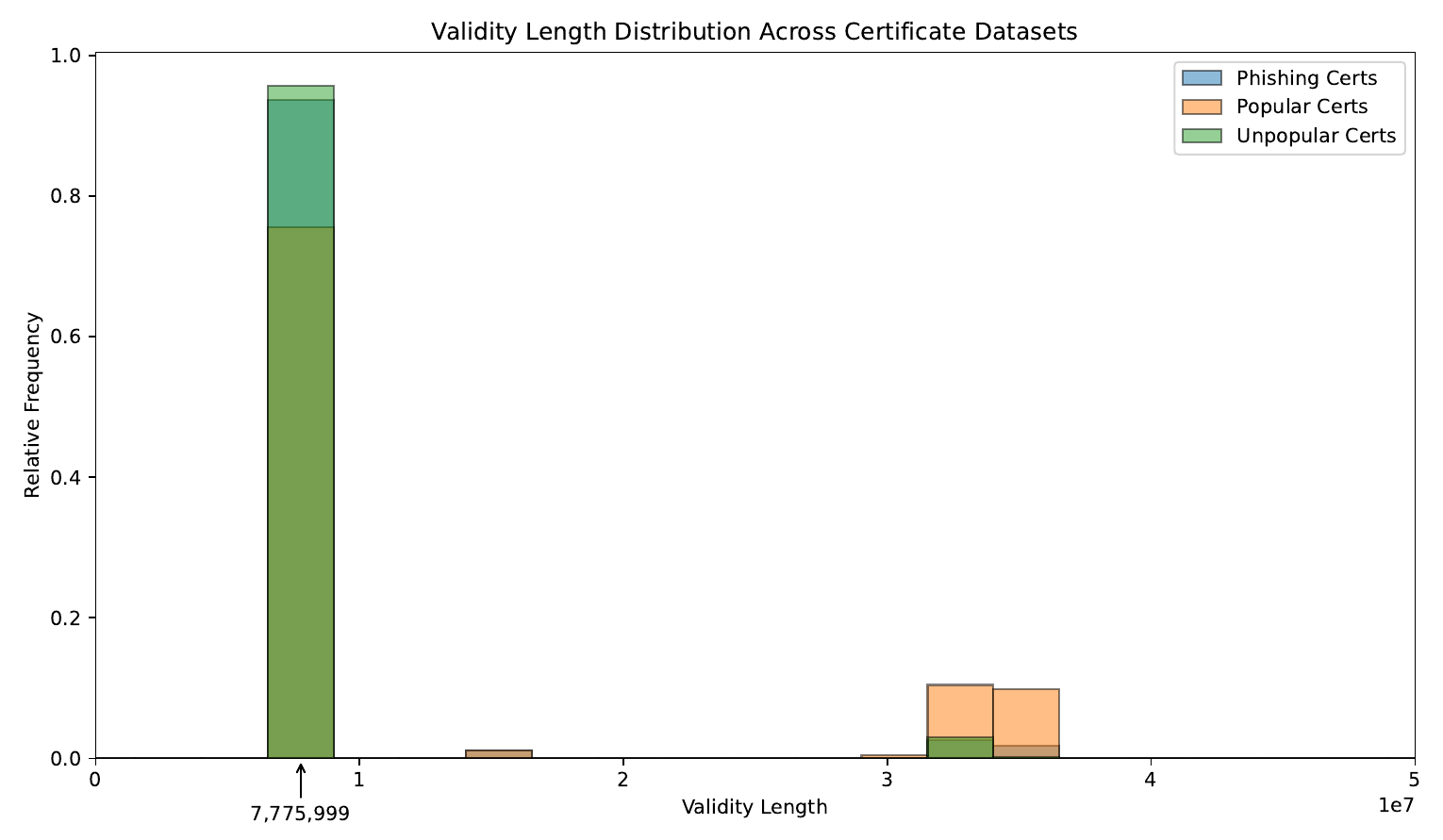}
    \caption{Histogram showing the relative frequency of certificate lifetimes for all datasets}
    \label{fig:histogram_relative_freq}
\end{figure}

Figure \ref{fig:histogram_relative_freq} shows that almost all certificate lifetimes within the unpopular and phishing datasets land in the same bin. Specifically $\sim$94\% of phishing certificates and $\sim$95\% of unpopular certificates. For the popular domains this is a bit lower, though still a large majority with $\sim$76\%, with most of the remaining certificates being in the two bins around 34,000,000 certificate lifetime. This diagram also shows that certificates issued for phishing domains and unpopular domains follow the same pattern. Based on this, it's not possible to identify whether or not a certificate was registered to a phishing domain based on certificate lifetime.

The hypothesis stated at the start of this analysis point was that certificates for phishing domains tend to have a shorter validity period than certificates for non-phishing domains. When looking at the validity durations of each dataset, only a single value stands out: the mean for validity length for popular certificates. While this value is higher, this is due to the higher population of domains utilizing certificates with a much longer lifespan. This means that if a certificate is registered with a longer lifetime, there is a greater chance that it wasn't registered with the intent of phishing. However, when looking at validity lengths in general, there are no feasible differences between certificates issued to phishing and benign domains in Denmark. 

\newpage

\subsection{Missing Fields}
This hypothesis examines whether missing fields in TLS certificates can indicate potential phishing activity. The idea is that phishing domains may use certificates that are less complete compared to those of legitimate and well-maintained domains. By examining both populated and missing certificate fields, we can analyze whether incomplete certificates are more common in phishing-related domains.

For each dataset category (\textit{phishing}, \textit{popular}, and \textit{unpopular}), the corresponding domains and their certificates were retrieved. For every certificate, the \textit{domain name}, the \textit{number of missing fields}, and the list of all \textit{missing fields} were collected. This allowed a systematic comparison of certificate completeness across the three groups.

To evaluate this hypothesis, descriptive statistics were calculated for each category. For every certificate group, the mean, median, and standard deviation of missing fields were computed, along with the minimum and maximum values.

\begin{figure}[h]
    \centering
    \includegraphics[width=0.8\linewidth]{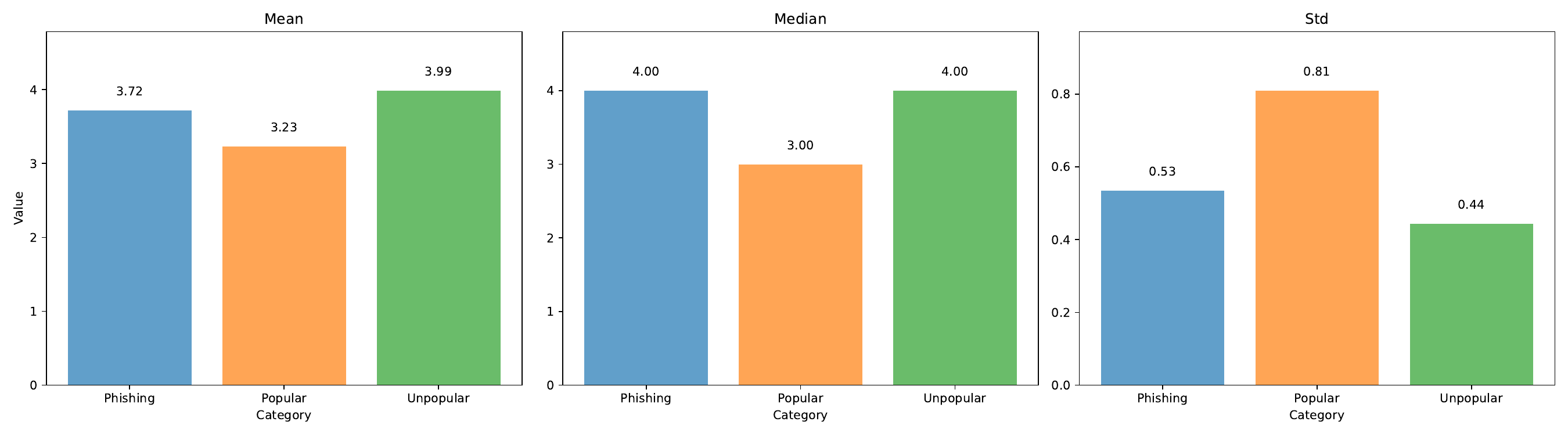}
    \caption{Main statistics of missing certificate fields across domain categories.}
    \label{fig:hyp3_main_stats}
\end{figure}
As shown in Figure~\ref{fig:hyp3_main_stats}, \textit{unpopular} domains have the highest average number of missing fields. The relatively low standard deviation indicates that this pattern is consistent across most certificates in this category.
\textit{Phishing} domains also show a relatively high average number of missing fields, with slightly more variation than the unpopular group. However, phishing certificates do not present extreme cases, as the maximum number of missing fields observed in this category is five. In contrast, popular domains have the lowest average number of missing fields, meaning that their certificates are generally more complete. However, this group shows the highest variation.

The results suggest that unpopular and phishing domains often have more incomplete certificates, while popular domains are generally more complete, indicating that missing certificate fields could serve as a feature for phishing detection, though dataset limitations and configuration differences may also influence these findings.

\begin{figure}[h]
    \centering
    \includegraphics[width=0.8\linewidth]{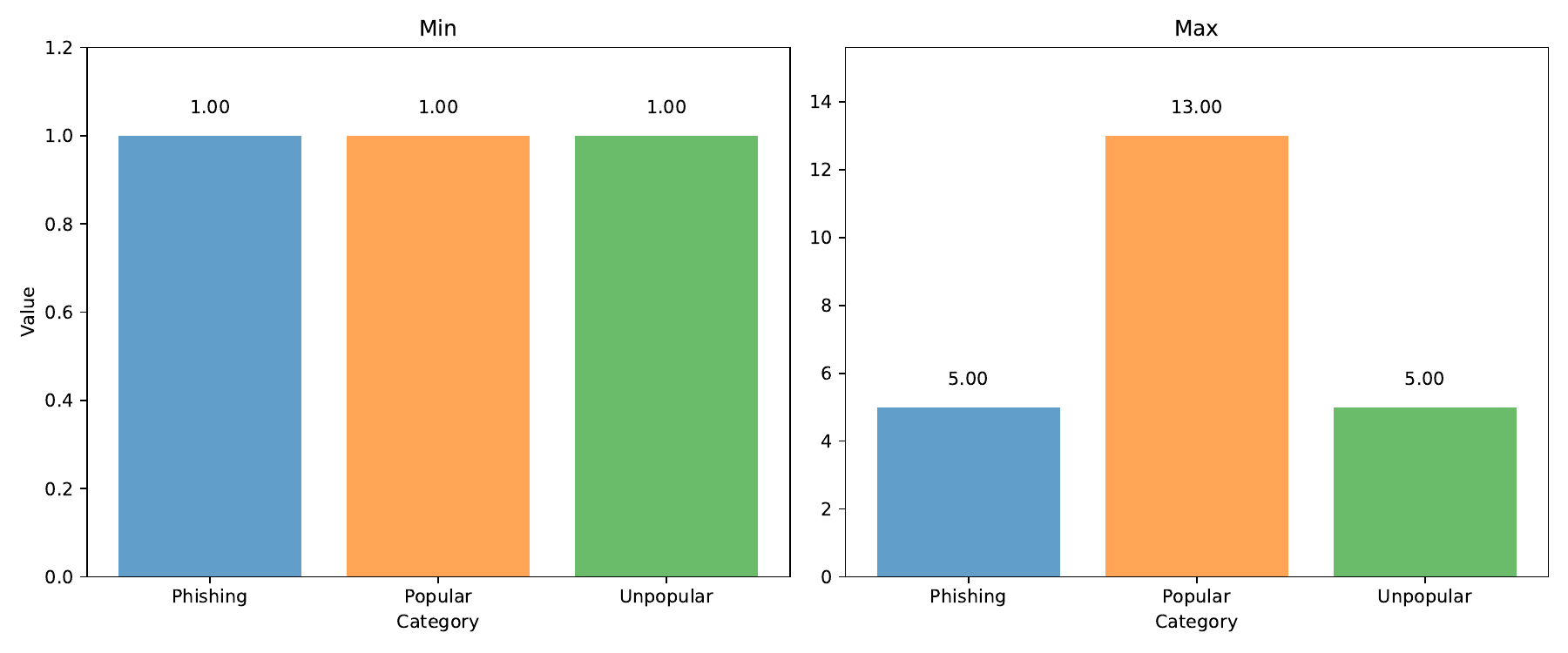}
    \caption{Minimum and maximum number of missing fields per certificate.}
    \label{fig:hyp3_minmax}
\end{figure}

Figure~\ref{fig:hyp3_minmax} shows that most certificates in the \textit{popular} category are complete, though a few have up to 13 missing fields, far exceeding the other categories. \textit{Phishing} domains also exhibit a relatively high number of missing fields with slightly more variation than the unpopular group, but none reach the extremes seen in popular domains. These patterns suggest that unpopular and phishing domains generally have more consistently incomplete certificates, while popular domains are mostly complete, a trend that may reflect differences in certificate quality or configuration practices, though dataset limitations could also play a role.

\begin{figure}[h]
    \centering
    \includegraphics[width=0.9\linewidth]{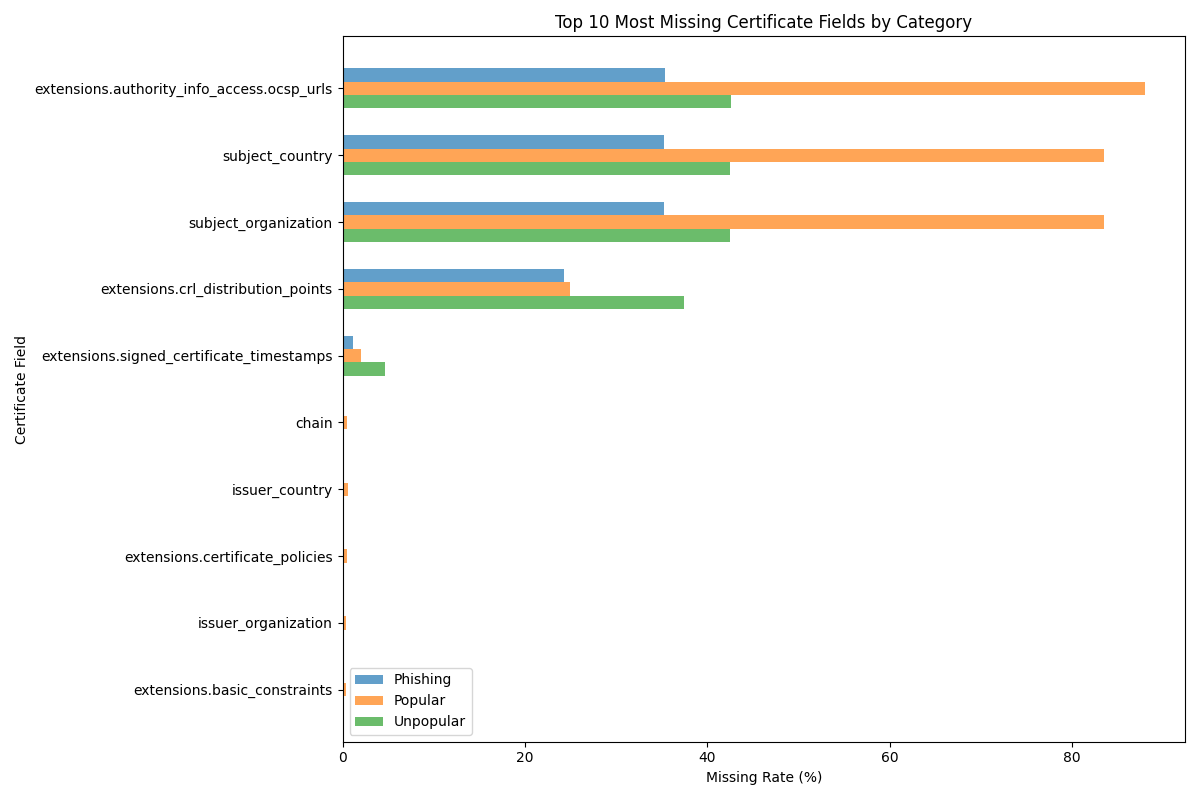}
    \caption{Top 10 Missing Fields by Category}
    \label{fig:top10_miss_fields_per_cat}
\end{figure}

Figure~\ref{fig:top10_miss_fields_per_cat} shows the ten most frequently missing fields per domain category. Fields like \textit{extensions.authority\_info\_access.ocsp\_urls}, \textit{subject\_organization}, and \textit{subject\_country} are commonly absent, reflecting typical DV certificate characteristics rather than malicious behavior. Other missing fields, such as \textit{extensions.crl\_distribution\_points} and \textit{extensions.signed\_certificate\_timestamps}, result from CA implementation differences and reliance on OCSP over CRLs. 

Overall, missing field patterns largely follow standard DV and CA practices, but disproportionately missing fields in phishing domains could still aid detection when combined with other features.

\subsection{SAN Similarity}\label{ch:san_analysis}
This hypothesis examines whether the Subject Alternative Names (SANs) in certificates associated with phishing domains are less internally similar than those found in benign certificates. Legitimate services typically manage their domains in a stable and organized manner, causing SAN entries to follow consistent naming patterns, whereas phishing infrastructure may be more heterogeneous and less structured.

Certificates from the three datasets were analyzed at the certificate level by extracting DNS SAN entries. Certificates containing zero or one SAN entry were excluded from the similarity analysis, as similarity cannot be computed in these cases. Internal SAN similarity was computed per certificate, and additional structural SAN characteristics were examined, including the total number of SAN entries, the presence of wildcard SANs, the number of distinct base domains, and the average depth of SAN labels, to provide contextual insight into certificate organization. Similarity was evaluated using multiple string-based similarity metrics, and consistent trends were observed across metrics.

\begin{figure}[h]
     \centering
     \includegraphics[width=0.95\linewidth]{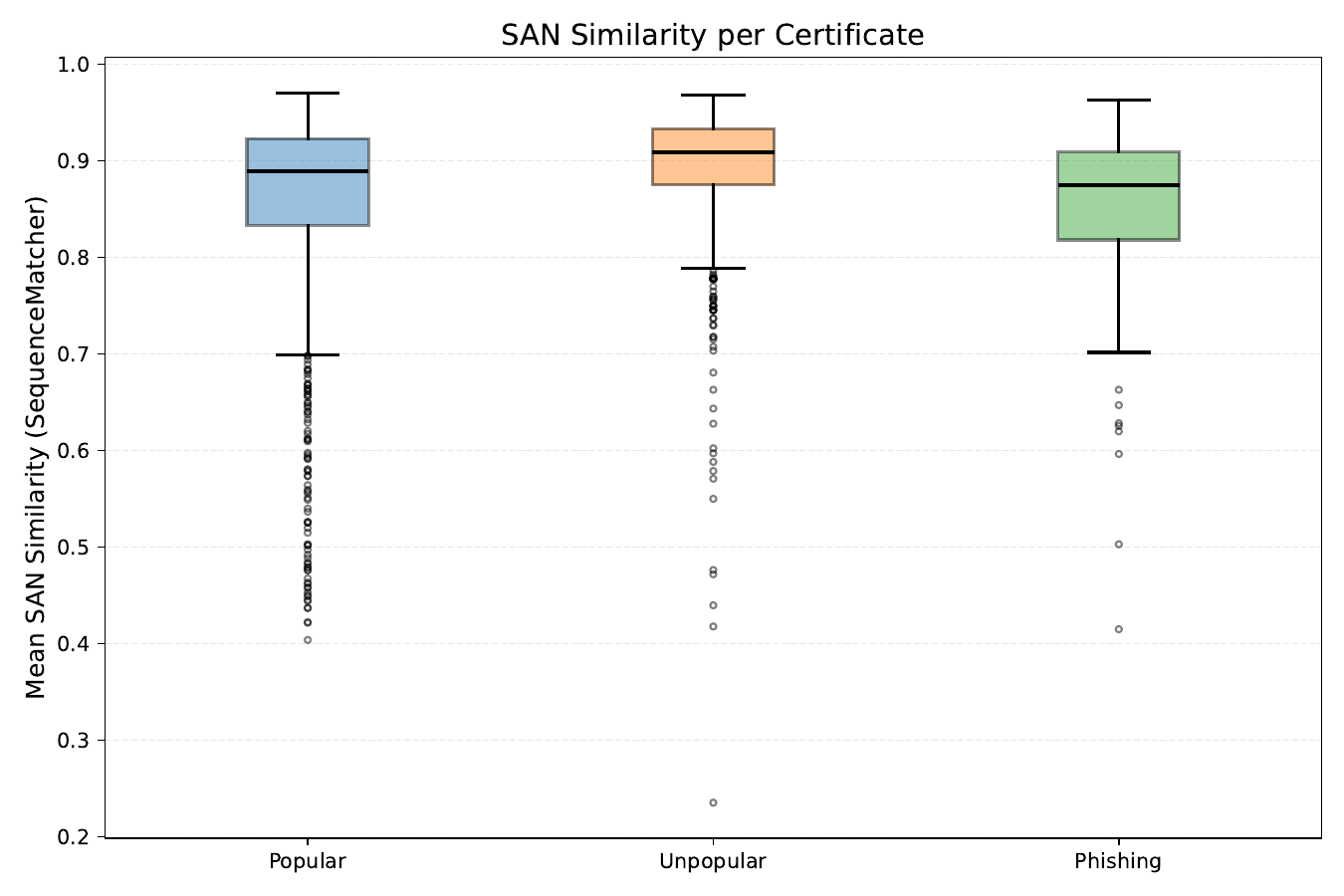}
     \caption{SAN Similarity per Certificate}
     \label{fig:h4-boxplot}
 \end{figure}

Figure ~\ref{fig:h4-boxplot} shows the distribution of SAN similarity scores across phishing, popular, and unpopular certificates. Phishing certificates tend to exhibit lower internal SAN similarity than certificates from popular domains, indicating less consistent naming structures. However, substantial overlap exists between phishing and unpopular domains. A Mann–Whitney U test indicates that the difference in SAN similarity between phishing and popular certificates is statistically significant, whereas no statistically significant difference is observed between phishing and unpopular certificates.
  
 \newpage
To further contextualize these findings, structural SAN characteristics are presented in Table~\ref{tab:structural_features}. Phishing certificates more frequently cover a diverse set of domains, while popular certificates tend to exhibit more structured and repetitive SAN patterns. Unpopular domains display behavior closer to phishing domains, reinforcing the observed overlap in similarity distributions.
\begin{table}[H]
\centering
\caption{Structural characteristics of SAN fields across certificate groups.}
\begin{tabular}{llll}
\toprule
\textbf{Feature} & \textbf{Phishing} & \textbf{Popular} & \textbf{Unpopular} \\ \midrule
Number of SANs (mean) & 0.80 & 3.01 & 0.99 \\
Unique base domains (mean) & 0.37 & 1.42 & 0.45 \\
Wildcard count (mean) & 0.10 & 0.49 & 0.23 \\
Average label count (mean) & 2.41 & 2.58 & 2.49 \\
\bottomrule
\end{tabular}
\label{tab:structural_features}
\end{table}
Overall, the results show that phishing certificates generally exhibit lower SAN similarity compared to certificates from popular domains, but SAN similarity alone does not provide a strong standalone indicator for phishing detection within the Danish namespace.

\subsection{Registrant country}

\begin{figure}[h]
    \centering
    \includegraphics[width=1\linewidth]{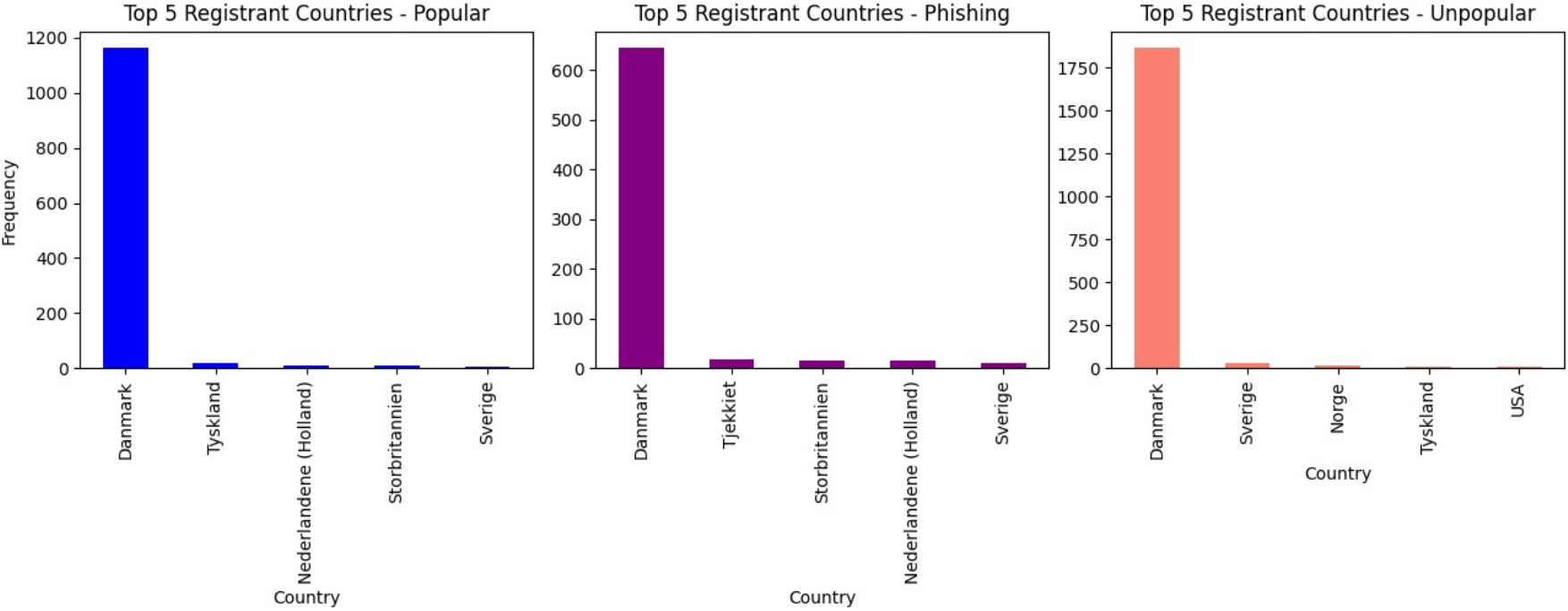}
    \caption{Distribution of registrant countries}
    \label{fig:regcountries}
\end{figure}

Many phishing domains are registered from organizations or individuals based outside of Denmark.
Moreover the registrant country distribution phishing, popular, and unpopular domains was compared to assess whether phishing domains are mainly registered outside Denmark. The distribution is visualized in Figure \ref{fig:regcountries}.

About 92\% of the popular .dk domain are registered within Denmark, likely because legitimate organizations maintain a local presence. Similarly, around 82\% of phishing domains and 90\% unpopular domains are also registered in Denmark.

Furthermore, the countries Sweden, Germany, Norway, the Netherlands, and the United Kingdom represent a small portion, registering 2-3\% of benign domains. While this number is not that large, it does indicate a trend to cross-border interest from neighboring countries, possibly to target Danish audiences. The distribution of phishing domains is slightly different from popular and unpopular ones, because Czechia appears among the top countries. Although the difference in the distribution between phishing and benign .dk domains is not large, the selection of the country Czechia may illustrate a trend for malicious actors.

The hypothesis, that most phishing domains are registered from outside Denmark, %regarding the registrant country of phishing domains
turns out to be not completely correct. Although in the given dataset around 18\% of the phishing domains are registered from countries outside Denmark, the majority of both subsets are registered domestically. This suggests that local actors may play a significant role in phishing activities targeting Danish domains. However, it cannot be ruled out the possibility that an attacker may use VPNs or other location-masking techniques to reduce suspicion and detection and increase trustworthiness.

\subsection{Phishers target specific sectors in Denmark for phishing impersonation}

This hypothesis proposes that phishers are not registering domains randomly, but are instead systematically targeting specific sectors in Denmark. 
The phishing domains was analyzed and assigned to one of several categories based on its name, structure, and likely impersonated target. The sector categories were:

\begin{itemize}
\itemsep0em 
\item Local Businesses, that consist of small to medium sized businesses that focus on provision of services or production, like trades and construction professionals, factories, farmers, beauty salons and technicians.  
\item Retail \& E-commerce, that includes every store that sells goods and not craftsmanship, either physically or online, from small local supermarkets to big retail companies 
\item IT \& Digital Services, that covers tech companies, web services and hosting infrastructure, software development, IT consult and support agencies, and famous tech brands.
\item Organizations (Public \& Private), consisting of educational institutions, non-profit clubs and associations, unions and foundations, community and cultural institutions, government infrastructure. 
\item Personal websites, which includes portfolios, personal blogs and journals, domains based on danish names or test websites created for fun.
\item Hospitality \& Wellness, incorporating restaurants, cafes, bars, gyms, massage clinics, personal coaches, hotels, sports and leisure establishments.
\item Healthcare, including medical providers and manufacturers, clinics, hospitals, pharmacies, health related services.
\item Finance, a sector that covers banks and financial institutions, insurance companies, accounting firms, payment methods.
\item Real Estate \& Housing, containing rental portals, property listings, real estate agencies, property managers, building associations and designers.
\item Logistics \& Shipping, including postal and courier services, freight and transport companies, package tracking portals and delivery notifications.
\item Unknown, which is the fail over category that contains generic domain names that cannot be identified.
\end{itemize}

The final distribution of phishing domains across these categories is shown in Table~\ref{tab:sectors}.

\begin{table}[h]
\caption{Phishing domain distribution by sector}
\centering
\begin{tabular}{ll}
\toprule
\textbf{Sector} & \textbf{Number of Domains} \\
\midrule
Local Business & 140 \\
Retail / E-commerce & 113 \\
IT / Digital Services & 111 \\
Organizations (Public \& Private) & 95 \\
Unknown & 90 \\
Personal Websites & 63 \\
Hospitality \& Wellness & 61 \\
Healthcare & 29 \\
Finance & 25 \\
Real Estate \& Housing & 21 \\
Logistics \& Shipping & 14 \\
\bottomrule
\end{tabular}
\label{tab:sectors}
\end{table}

The 5 most targeted sectors were local businesses with 140 domains, retail \& E-commerce with 113 domains, IT \& digital services with 111 domains, organizations (public \& private) with 95 domains, and unknown with 90 domains. Together they sum to 549 domains, accounting for about 72\% of the dataset. Removing the domains with unknown target from this, brings the sum down to 459 domains, which accounts for around 60\%. The personal websites and hospitality \& wellness sectors account for respectively 63 and 61 domains. The final 4 sectors, healthcare, finance, real estate \& housing, and logistics \& shipping account for 29, 25, 21, and 14 domains. 

Four of the sectors account for the vast majority of certificates (60\%). This shows that phishers do not randomly select targets within the Danish namespace, but rather have a clear pattern of sectoral focus. Specifically, local businesses, retail \& E-commerce, IT \& digital services, and organizations (public \& private) are the most targeted. Therefore, this hypothesis is supported.

\subsection{Structural and lexical domain-based features}

\begin{figure}[h]
    \centering
    \includegraphics[width=1\linewidth]{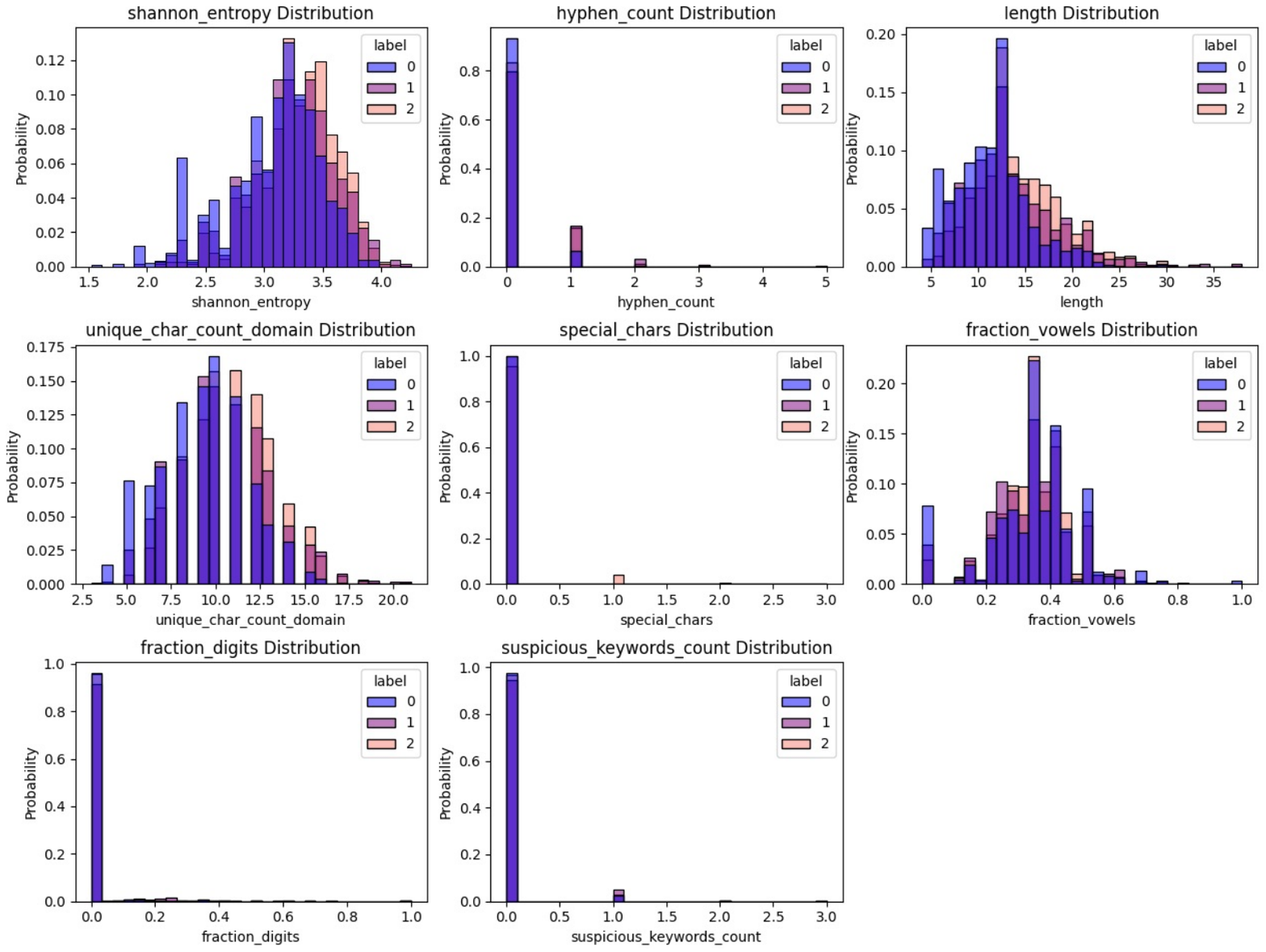}
    \caption{Probability distribution of numerical domain-based features for each label}
    \label{fig:prob_dist}
\end{figure}

This section presents the descriptive analysis of domain-based features across popular, unpopular, and phishing domains. 
Although most of the investigated features did not show huge differences, certain characteristics revealed notable trends that could contribute to improving detection of possible phishing domains. The results are visualized using histograms in Figure \ref{fig:prob_dist} (label 0: popular, label 1: phishing, label 2: unpopular).

One of the key findings was that phishing domains and unpopular domains tend to have a slightly longer domain name compared to popular domains. However, the difference in domain length was minimal and, therefore, not a strong standalone indicator of phishing activities. Shannon Entropy, which measures randomness of an input, shows slightly higher values for phishing and unpopular domains, but no clear separation between categories. A higher value correlates to a more random or unpredictable input.

Moreover, the hyphen count was analyzed to investigate whether it differs between phishing and popular domains. While most domains across all 3 datasets contain zero hyphens, about 20\% of the phishing domains have one hyphen compared to roughly 5\% of popular domains. However, there is a overlap between phishing domains and unpopular domains. Similarly, phishing and unpopular domains tend to use slightly more unique characters, but the difference remains subtle.

Other features, including the count of special characters, the portion of vowels in the domain name, the occurrence of suspicious keywords based on different lists ~\cite{gareraFrameworkDetectionMeasurement2007,drichelFindingPhishHaystack2021}, and the portion of digits, show no clear separation between categories. The results show that these features cannot be used to differ between phishing and benign .dk domains. Additionally, no punycode-encoded domains or homoglyphs were detected, suggesting that such IDN-based obfuscation techniques are not commonly used in this dataset.

\newpage
\subsection{Choice of hosting providers}

\begin{figure}[h]
    \centering
    \includegraphics[width=1\linewidth]{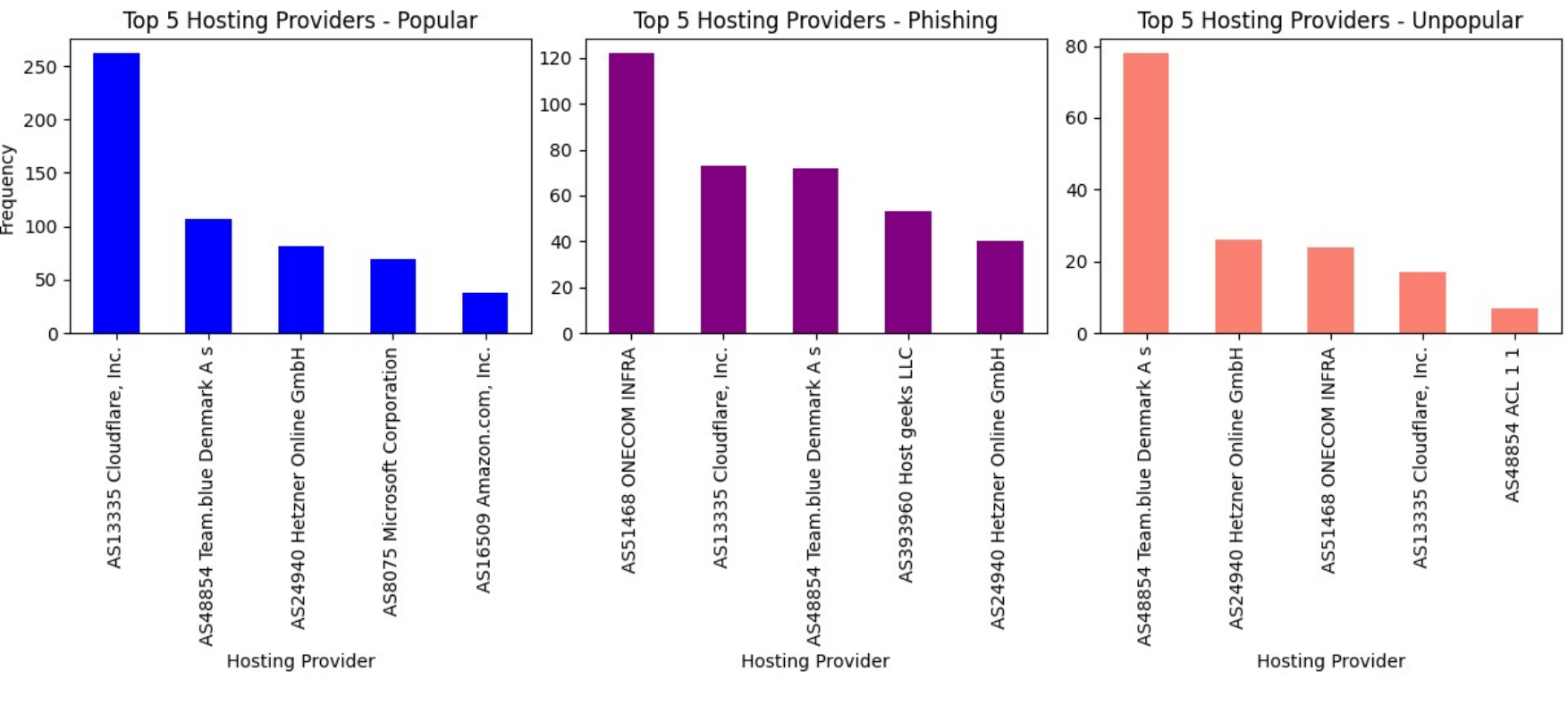}
    \caption{Distribution of hosting providers}
    \label{fig:hostprovider}
\end{figure}

To investigate whether phishing and benign domains differ in their choice of hosting providers, these were extracted using ipwhois \footnote{IPWhois is a service for querying and parsing WHOIS and RDAP data of IP addresses to obtain information about network ownership and registration details. \url{https://ipwhois.io/}}. The analysis focuses on the organizations responsible for hosting the resolved IP addresses of the domains. The distribution of the five most frequent hosting providers for each label is illustrated in Figure \ref{fig:hostprovider}. 

No clear distinction in the choice of hosting providers was observed between resolvable phishing, popular, and unpopular domains in the datasets. Some providers like \textit{Host geeks LLC} only occurred for phishing domains, but the overall provider distribution still overlaps with popular and unpopular domains. This shows that phishing actors do not rely on specific providers. Therefore, the lack of variation in the dataset suggests that the choice of the hosting provider alone is not a reliable indicator of phishing activity.

\section{Discussion and future work}
\label{sec:discussion}

Several challenges were encountered during the collection and analysis of the data, which are discussed in this section.

Apparent from the beginning was that not all domains would have available certificates. However, this turned out to be a larger problem than initially estimated. Initially, OpenIntel and Censys were considered as potential CT log data sources. However, due to access limitations, the Netlas threat intelligence platform was chosen instead. This data is, however, not live data from CT log infrastructure. In the end only 35\%, 42\% and 88\% of certificates were collected from phishing, unpopular, and popular domains respectively. It is likely that the problem is with Netlas as a data source. Since the results of the analysis don't differ too much from results from related work, the collected data is still representative of phishing strategies as a whole. %Ideally, though, with better access to CT logs more certificates could have been retrieved, resulting in a more precise analysis.

Moreover, because the data is historic, one domain may throughout it's lifetime have had multiple certificates. It is especially problematic in the cases that a domain has changed ownership, and may at one point be benign and at another phishing. Therefore, it was important to retrieve one appropriate certificate for each domain.

Another limitation of this study is that the analysis primarily considers individual features in isolation. For each hypothesis, individual features were analyzed independently, without considering potential interactions between multiple features. This does not capture potential interaction or correlations between features, which may jointly improve discrimination between phishing and non-phishing domains. 

Future work should therefore investigate multi-variable analysis, for example by exploring correlations between features or applying machine learning techniques to evaluate their combined predictive power. Moreover, a field of analysis that is missing from this project is phishing utilizing subdomains. Further analysis of phishing activity involving subdomains may reveal additional patterns in attacker behavior and infrastructure usage. Finally, exploring differences and similarities across spam, phishing, and malware domains could clarify whether the observed certificate and domain characteristics are phishing-specific or represent general indicators of malicious behavior.

% not alle domains had certificate -> because of Netlas
% one certificate 
% single-variable analysis -> future work: multi-variable analysis in machine learning

% future work: other malicious types, subdomains

\section{Conclusion}
\label{sec:conclusion}
In this study a quantitative analysis of digital certificates and domain characteristics was presented to determine whether phishing domains can be distinguished from benign domains in the Danish namespace. Using a dataset consisting of phishing, popular and unpopular domains, certificates were retrieved through
Netlas.

The analysis shows that across multiple metrics, phishing domains in the
Danish namespace differ from popular domains in several aspects, but these differences are generally much smaller between phishing and unpopular. Both certificate-based and domain-based features were examined to assess their ability to distinguish phishing domains from benign domains. While certain tendencies were observed, the differences were generally small and showed significant overlap, especially when comparing phishing and unpopular domains.

In summary, this study concludes that no single analyzed feature provides a strong standalone indication of phishing activity within the Danish namespace.
While phishing domains exhibit some tendencies, such as lower SAN similarity or longer domain names, these patterns are weak and insufficient on their own. However, these tendencies could be leveraged in more sophisticated detection mechanism.

\section*{Acknowledgments}

The authors thank Punktum dk for providing access to registry data and phishing reports used in this study. We also acknowledge Netlas for access to TLS certificate data used in the analysis. In addition, we thank Censys for granting research access to their platform, although we did not use Censys due to time constrains.

%Bibliography
\bibliographystyle{unsrt}  
\bibliography{mybib}

\end{document}